\documentclass{emulateapj}

\usepackage{natbib}    
\usepackage{graphicx,float}
\usepackage{rotating}

\begin{document}

\title{HYPERVELOCITY STAR CANDIDATES IN THE SEGUE G \& K DWARF SAMPLE}

\author{Lauren E. Palladino\altaffilmark{1}, Katharine J. Schlesinger\altaffilmark{2}, Kelly Holley-Bockelmann\altaffilmark{1,3}, Carlos Allende Prieto\altaffilmark{4,5}, Timothy C. Beers\altaffilmark{6}, Young Sun Lee\altaffilmark{7}, Donald P. Schneider\altaffilmark{8,9}}

 \affil{
     $^{1}$ Department of Physics and Astronomy, Vanderbilt University, Nashville, TN 37235, USA\\
     $^{2}$ Research School of Astronomy and Astrophysics, The Australian National University, Weston, ACT 2611, Australia\\
     $^{3}$ Department of Natural Sciences and Mathematics, Fisk University, Nashville, TN 37208, USA\\
     $^{4}$ Instituto de Astrof\'{\i}sica de Canarias, E-38205 La Laguna, Tenerife, Spain\\
     $^{5}$ Departamento de Astrof\'{\i}sica, Universidad de La Laguna, E-38206 La Laguna, Tenerife, Spain\\
     $^{6}$ National Optical Astronomy Observatory Tucson, AZ, 85719, USA\\
     $^{7}$ Department of Astronomy, New Mexico State University, Las Cruces, NM 88003, USA\\
     $^{8}$  Department of Astronomy and Astrophysics, The Pennsylvania State University, University Park, PA 16802, USA\\
     $^{9}$ Institute for Gravitation and the Cosmos, The Pennsylvania State University, University Park, PA 16802, USA\\
     lauren.e.palladino.1@vanderbilt.edu, k.holley@vanderbilt.edu}

\begin{abstract} 
We present 20 candidate hypervelocity stars from the Sloan Extension for Galactic Understanding and Exploration (SEGUE) G and K dwarf samples. Previous searches for hypervelocity stars have only focused on large radial velocities; in this study we also use proper motions to select the candidates. We determine the hypervelocity likelihood of each candidate by means of Monte Carlo simulations, considering the significant errors often associated with high proper motion stars. We find that nearly half of the candidates exceed their escape velocities with at least 98\% probability. Every candidate also has less than a 25\% chance of being a high-velocity fluke within the SEGUE sample. Based on orbits calculated using the observed six-dimensional positions and velocities, few, if any, of these candidates originate from the Galactic center. If these candidates are truly hypervelocity stars, they were not ejected by interactions with the Milky Way's supermassive black hole. This calls for a more serious examination of alternative hypervelocity-star ejection scenarios.
\end{abstract}


\section{INTRODUCTION}

Hypervelocity stars (HVSs) are believed to be ejected by three-body interactions with the supermassive black hole (SMBH) at the Galactic center \citep[e.g.,][]{Hills,Yu,Brown2005}.  During this process, energy and angular momentum are transferred from the black hole to one of the stars in a binary system.  The second star loses energy and becomes bound to the black hole while the first  is ejected from the Galaxy. In this scenario HVSs can probe conditions in the Galactic center such as the binary fraction, and even place limits on the existence of a second, tightly bound SMBH. Semianalytical models predict that there may be approximately 100 HVSs within 8 kpc of the Galactic center due to the break up of equal-mass binaries \citep{GouldQuillen2003,Yu}.

While the SMBH at the Galactic center remains the most promising culprit in generating HVSs, other hypervelocity ejection scenarios are possible, such as a close encounter of a single star with a binary black hole \citep{Yu}. In this case, the star gains energy from the binary black hole and is flung out of the Galaxy while the orbit of the black hole binary shrinks \citep[e.g.,][]{Quinlan,Sesana}. Another alternative hypervelocity ejection model involves the disruption of a stellar binary in the Galactic disk; here a supernova explosion in the more massive component can accelerate the companion to hypervelocities \citep[e.g.,][]{Blaauw1961,LeonardDewey1993,Napiwotzki2012}. 

At least 18 HVSs have been discovered in the Milky Way within the past decade with velocities as high as 700 km s$^{-1}$ \citep[e.g.,][]{Brown2005,Edelmann2005,Hirsch2005,Brown2009,BrownGeller2012}. So far, all confirmed HVSs are massive B-type stars such as those observed around the central SMBH \citep[e.g.,][]{Brown2009,BrownGeller2012}. However, since the ejection mechanisms described above apply to any stellar mass, it is important to search for HVSs among the larger set of longer-lived, lower mass stars \citep[e.g.,][]{Quinlan}. If a SMBH ejection mechanism is at play, then metal-rich stars originating from the Galactic center ought to pollute the metal-poor halo. Previous attempts to mine Sloan Extension for Galactic Understanding and Exploration (SEGUE) and SEGUE-2 stellar halo data found no metal-rich, old ejected stars \citep{Kollmeier2009,Kollmeier2010}. The lack of a significant population of old, metal-rich HVSs suggests that the initial mass function at the Galactic center is mildly top-heavy. Alternatively, hypervelocity ejection mechanisms may be more complex than previously thought.

In this paper we identify the first set of G- and K-type candidateHVSs from SEGUE. We discuss candidate selection in Section 2, including a description of the G and K dwarf sample,  and we address the significant proper-motion errors in Section 3. Section 4 contains orbital parameters for the HVS candidates, and Section 5 discusses possible alternative origin scenarios. Finally, we summarize and conclude in Section 6.

\begin{figure}
\includegraphics[width=0.5\textwidth]{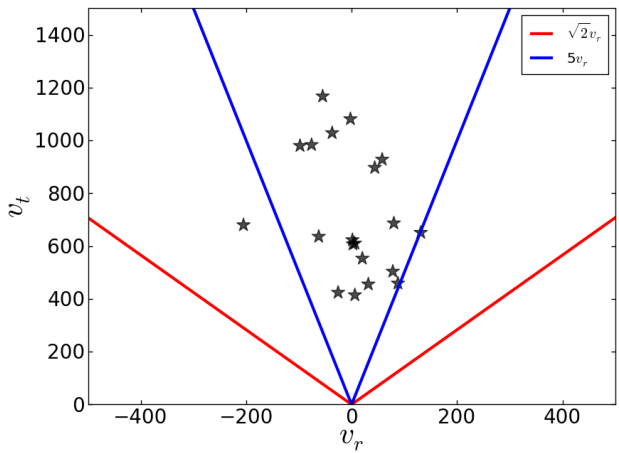}
\caption[\,\,]{\footnotesize Transverse versus radial velocities of our HVS candidates, in kilometers per second. Red lines indicate a transverse velocity $\sqrt{2}$ times higher than the radial velocity, as expected for an isotropic stellar distribution. Blue lines represent a transverse velocity 5 times higher than the radial velocity. The majority of our candidates show large transverse-to-radial velocity ratios, characteristic of a sample strongly affected by large proper-motion errors. We caution that some of our HVS candidates may be high-velocity flukes, and we calculate the likelihood of this in Section 3.2.}
\label{}
\end{figure}

\begin{figure*}[]
\centering
\begin{tabular}{cc}
\includegraphics[width=6.6in]{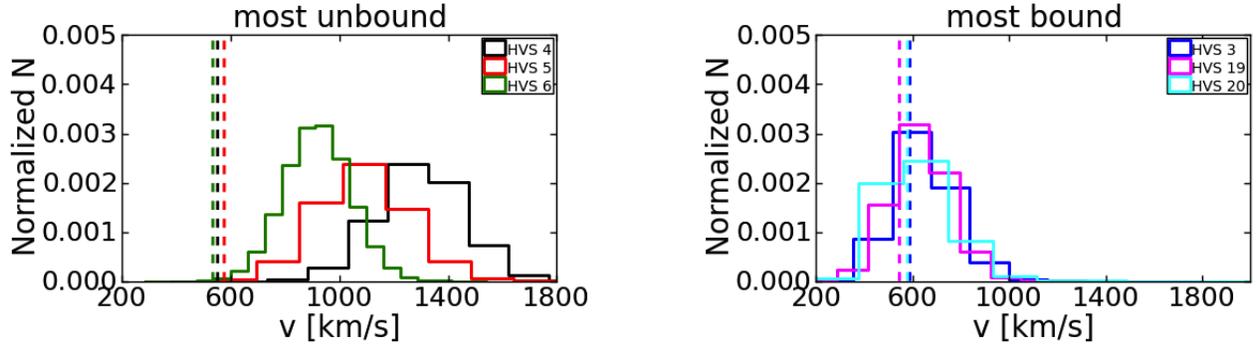} 
\end{tabular}
\caption[\, \,]{\footnotesize Velocity distribution for a million random samples of the velocity error distribution for the three least and most bound HVS candidates. Dashed lines show escape velocity of each candidate.}
\label{}
\end{figure*}

\section{Identifying HVS Candidates}


Our candidates are drawn from the G and K dwarf stars in SEGUE \citep[][]{Yanny2009} from the Sloan Digital Sky Survey (SDSS) Data Release 9 \citep[DR9;][]{DR9}. As part of SDSS \citep[][]{York2000}, SEGUE provides medium-resolution ($R\approx$ 1800) spectroscopy over a broad spectral range (3800--9200 \AA). Probing more than 150 lines of sight, SEGUE covers $\approx$3500 deg$^2$ of the sky, with spectroscopy of  $\approx$240,000 stars over a range of spectral types. Technical information about SDSS has been published on the survey design \citep[][]{York2000,Eisenstein2011}, telescope and camera \citep{Gunn98,Gunn2006}, and spectrographs \citep{Smee2012}, as well as the photometric system \citep{Fukugita96} and astrometric \citep{Pier2003} and photometric \citep{Ivezic2004} accuracy.

G and K dwarfs are selected from the SDSS photometric data using simple color and magnitude selection criteria. The 42,901 SEGUE G dwarfs are defined as having 14.0 $< r_0 <$ 20.2 and 0.48 $< (g-r)_0 <$ 0.55, while the 28,332 K dwarfs have 14.5 $< r_0 <$ 19.0 with 0.55 $< (g-r)_0 <$ 0.75 \citep{Yanny2009}. The subscript zero indicates that the color and magnitude have been corrected for dust extinction, using estimates derived from \cite{Schlegel}. Each spectrum is analyzed with the DR9 SEGUE Stellar Parameter Pipeline (SSPP), which provides estimates of effective temperature, surface gravity ($\log g$), [Fe/H], and [$\alpha$/Fe] \citep{Lee2008I,Lee2008II,Carlos2008,Smolinski2011,Lee2011}. We follow the quality protocol of \cite{Schlesinger2012} to remove targets with poor signal-to-noise ratio (S/N $< $10), incalculable atmospheric parameters, excessive reddening (greater than 0.5 mag in $r$),  saturated photometry ($r_0$ $< $ 15), or flags indicating temperature or noise issues. We also use the SSPP $\log g$ estimates to ensure the stars are dwarfs, using a cut on $\log g$ as a function of [Fe/H] to isolate dwarf stars (K. J. Schlesinger et al. 2014, in preparation).

For each star that satisfies these criteria, we determine its distance using the isochrone-matching technique described by \cite{Schlesinger2012}. Briefly, each star is matched in [Fe/H] and $(g-r)_{0}$ to 10 Gyr isochrones from the empirically corrected Yale Rotating Stellar Evolution Code set \citep{An2009}. There are systematic distance uncertainties introduced by using 10 Gyr isochrones, as well as the possibility of undetected binarity; this leads to a systematic shift in distance of $-$3$\%$ for the most metal-rich stars, while metal-poor stars are largely unaffected; this is factored into our distance estimates. There are also random distance errors from uncertainties in photometry, [Fe/H], [$\alpha$/Fe], and, finally, isochrone choice. The total random distance uncertainty is dominated by uncertainties in [Fe/H] and ranges from around 18$\%$ for stars with [Fe/H] $> -$0.5$\%$ to 8$\%$ for more metal-poor stars. 

To identify HVSs, we convert the radial and tangential velocities to Galactic Cartesian coordinates, as described below in Section 4.1, and choose a simple but conservative initial total velocity threshold of 600 km s$^{-1}$ to identify stars that exceed the Galaxy's escape velocity \citep[e.g.;][]{Smith}. We then verify that each candidate exceeds the escape velocity at its current location within the Galaxy. This procedure yields 42 preliminary HVS candidates of varying quality, which we further glean as described in Section 3.1 below.

\section{Estimating the Fidelity of Our Candidates}
\subsection{Proper-Motion Quality Cuts}

The proper-motion distribution in SDSS is skewed toward large proper-motion errors. We must ensure that the extreme velocities of our candidates are real rather than the product of large errors.
We describe our technique to ensure the robustness of our candidates in this section. The first step in defining a clean hypervelocity sample is to assess the quality of the proper-motion measurement. To determine the proper motions for SDSS targets, \cite{Munn2004,Munn2008} matched each SDSS point source to the USNO-B catalog. The resulting SDSS+USNO-B catalog is 90$\%$ complete to $g <$19.7 and has statistical errors of approximately 3$-$3.5 mas yr$^{-1}$ and systematic errors of $\approx$0.1 mas yr$^{-1}$ for each component. 

\cite{Munn2004} defined a number of criteria to ensure that the SDSS+USNO-B proper motions are reliable; these conditions resulted in a version of the USNO-B catalog with a contamination of less than 0.5$\%$. The criteria were later revised by \cite{Kilic2006} and are as follows\footnote{The parameters listed in parentheses are available in the DR9 proper motions catalog in the SDSS Catalog Archive Server (CAS).}: 
\begin{itemize}
\item The number of objects in USNO-B within a 1" radius of the SDSS target should be 1 ({\tt match$=$1}).
\item The rms residual for the proper-motion fit in right ascension and declination must be less than 525 mas ({\tt sigRA $<$ 525 and \tt sigDEC $<$ 525}).
\item There must be at least six detections (including the SDSS observations) used to determine the proper motion ({\tt nFit $=$ 6}).
\item The distance to the nearest neighbor with $g <$ 22 must be greater than 7" ({\tt dist22 $>$ 7}).
\end{itemize} 

Only three of our 42 preliminary candidates met all of the proper-motion quality criteria, and we categorize these as ``Clean." We performed an in-depth analysis for the remaining 39 stars, assigning each a likelihood of proper-motion contamination based on the same criteria adopted by \cite{Kilic2006} for their white dwarf sample. They found that the chance of contamination for a target with
\begin{itemize} 
\item Six detections and a neighbor within 7" is less than1.5$\%$;
\item Five detections and no near neighbors is 1.5$\%$;
\item Five detections and a neighbor within 7" is 35$\%$; 
\item Four detections and no neighbor within 7" is 51$\%$; and
\item Three detections and no neighbor within 7" is 89$\%$.
\end{itemize} 
We further checked for any potential blending issues by visually inspecting each candidate. 

We categorize 17 stars as having ``Reliable" proper motions, with 1.5\% or less chance of contamination and no visual blending. We categorize 10 stars as ``Possible," meaning they have between 35\% and 51\% chance of contamination. Twelve stars were removed because of visual blending.

We choose to consider only those candidates with ``Clean" and ``Reliable" designations. Thus, our final sample contains 20 HVS candidates, all with greater than 98.5$\%$ probability of robust proper-motion estimates.

We do expect that this final sample contains false-positive HVS detections. One way to illustrate this is with Figure 1, which compares the transverse and radial velocities of our HVS sample. For a random isotropic stellar distribution, the transverse velocity should be roughly $\sqrt{2}$ times higher than the radial velocity, and the fact that this sample is predominantly composed of stars with much larger transverse-to-radial velocity ratios is a classic signature of contamination by large proper-motion errors. While this does not prove that all of the candidates are spurious, it does indicate that many of them may be. Of course, a true hypervelocity sample would not be well represented by a random, isotropic distribution, but we caution that it is premature to say that we have identified 20 HVSs. We conducted further statistical tests, described below, to evaluate the likelihood that each HVS candidate is real.

\begin{figure*}[]
\centering
\begin{tabular}{cc}
\includegraphics[width=6.6in]{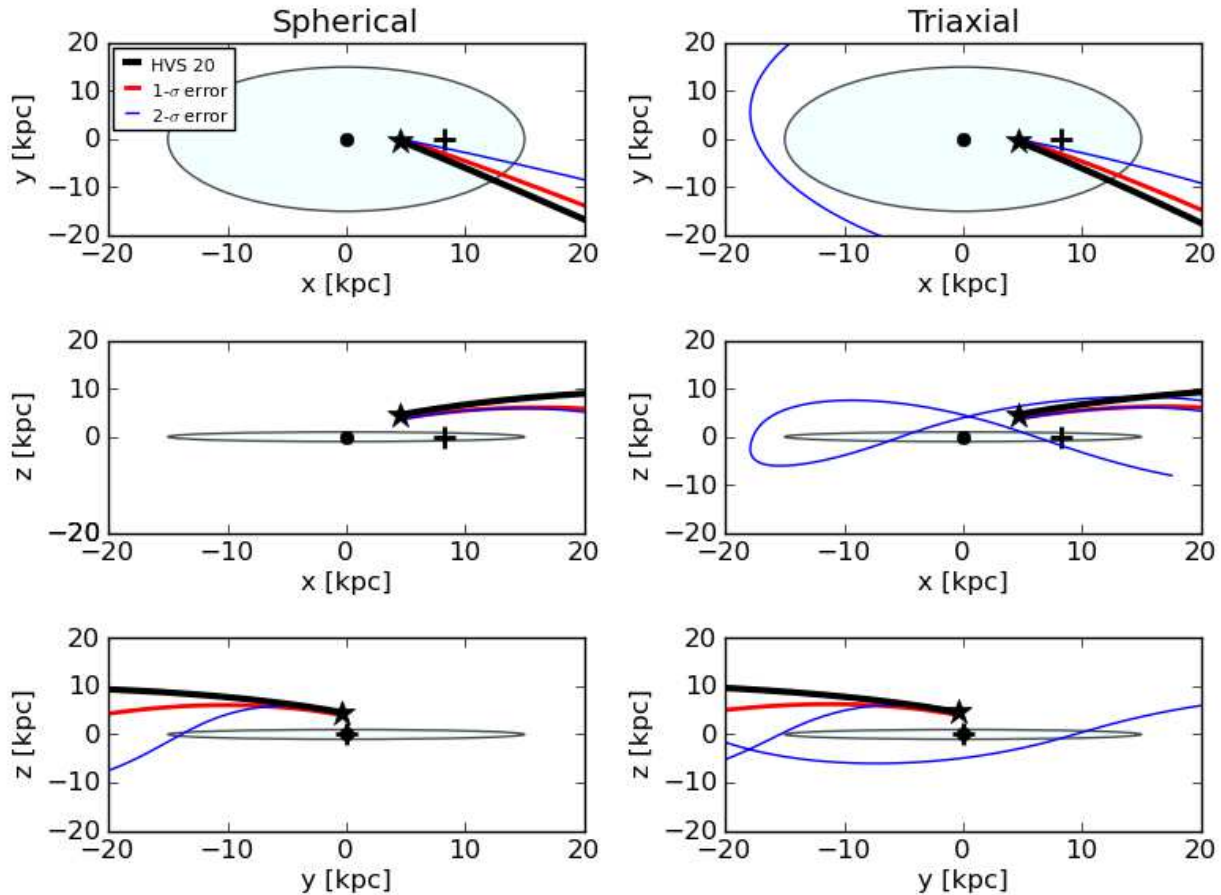} 
\end{tabular}
\caption[\, \,]{\footnotesize Orbit of HVS 20, a candidate with a ``Reliable" proper-motion measurement but the largest probability of being bound, shown by the black lines.  Also shown are the resulting orbits for the same candidate with 1$\sigma$ (red lines) and 2$\sigma$ (blue lines) velocity errors from the million Monte Carlo realizations. Left, two-dimensional projections in the spherical dark matter halo; right, the same for the triaxial model. The black dots and plus signs represent the locations of the Galactic center and the Sun, respectively, while the pale blue ellipses provide a rough scale for the extent of the disk. The five-pointed star in each panel marks the current position of HVS 20. The top row is a top-down view of the Galaxy while the middle and bottom rows are side views along the disk. Here we show that some candidates may in fact live on very bound orbits, and in such cases the shape of the orbit is strongly influenced by the triaxiality of the halo.}
\label{}
\end{figure*}

\subsection{Monte Carlo Sampling}

Although the typical error in proper motion in the SEGUE database is $\sim 10$ mas yr$^{-1}$, proper-motion errors can, for some stars, be much larger than expected for a normal distribution, especially at the high-velocity end \citep[][]{Gould2003,GouldSalim2003,Gould2004,Munn2004,Munn2008,Bond2010}. With this in mind, we consider the possibility that these HVS candidates may have true velocities much lower than can be explained by the reported errors and that they are in fact bound to the Galaxy. 

In order to determine the true range of velocities for our HVS candidates, as well as the probability that these candidates are bound given a more realistic error distribution, we built a Monte Carlo simulation to sample possible orbital parameters for each HVS candidate. \cite{Dong2011} obtained a proper-motion error distribution for the SDSS+USNO-B catalog by compiling proper motions for a sample of SDSS quasars that met the \cite{Kilic2006} criteria. We randomly resampled a million realizations of each HVS candidate's kinematics from the \cite{Dong2011} non-Gaussian proper-motion error distribution and Gaussian radial velocity errors. We also resampled each candidate's position, assuming Gaussian errors in the distance determinations as well. We find that 13 of the 20 candidates remain hypervelocity with greater than 90\% probability.

Figure 2 shows the distribution of velocities drawn randomly from the errors for the three least and most bound candidates. In most cases, the drawn velocity well exceeds the escape velocity, represented by the vertical dashed lines.

We performed a second Monte Carlo test to quantify the chance that these high velocities are simply the extreme tail end of the velocity error distribution within the entire SEGUE G and K dwarf sample. Here we construct a new mega-SEGUE sample built from 1000 realizations of each SEGUE star, in which each realization is drawn from the error distribution in proper motion, radial velocity, and distance as described above. We then calculate the ``interloper likelihood" for each candidate with respect to the mega-SEGUE sample; this is the probability that a slow, noncandidate star within our sample could have had the observed velocity of a particular candidate, given the errors:

\begin{equation}
P(\rm{interloper,i}) = 1 - \frac{\rm{n}_{HVS}(\rm{v} \ge \rm{v}_{\rm{cand},i})}{\rm{n}_{tot}(\rm{v} \ge \rm{v}_{\rm{cand},i})},
\end{equation}
where n$_{HVS}$($v$ \, $\ge$ \,  $v_{\rm{cand},i}$) is the number of stars in the mega-SEGUE sample with velocity greater than or equal to the observed velocity of candidate $i$ that were originally tagged as hypervelocity in the data and  n$_{tot}$($v$ \, $\ge$ \,  $v_{\rm{cand},i}$) is the total number of stars in the Monte Carlo sample with velocity greater than or equal to the candidate's velocity. All candidates have less than a 25\% ``interloper likelihood," and more than half have less than 10\%. Together, these two tests indicate that some of our candidates may in fact be the result of a statistical fluke. However, we expect the bulk of the candidates to remain hypervelocity.

Table 1 lists the velocity of each candidate determined from the proper motions reported in DR9, the minimum velocity calculated from a million realizations of the proper motion, radial velocity, and distance errors, the escape velocity for each candidate in a spherically symmetric Galaxy, the probability that the candidate may be bound given the escape velocity at its position, and the interloper likelihood as described above.

\section{Orbits of HVS Candidates}
\subsection{Galaxy Model}

We construct an analytical, multicomponent model of the Milky Way gravitational potential to predict the orbits of stars in the Galaxy based on the initial six-dimensional observed position and velocity. The model is easily modifiable and can be tuned to reflect the observed Galactic structural parameters. 

Our model includes the following components: a central SMBH with $M_{\rm{SMBH}} = 4 \times 10^6 M_{\odot}$; a spherical Hernquist bulge \citep{Hernquist1990} with $M_{\rm{bul}} = 4.5 \times 10^9 M_{\odot}$ and $r_{\rm{bul}} = 2.5$ kpc; Miyamoto$-$Nagai thin and thick disks \citep{MiyamotoNagai1975} with $M_{\rm{thin}} = 6 \times 10^{10} M_{\odot}, M_{\rm{thick}} = 6 \times 10^9 M_{\odot}$, 0.3 kpc thin-disk scale height, 1 kpc thick-disk scale height, and 3 kpc scale lengths for both; and a Navarro$-$Frenk$-$White (NFW) dark matter halo \citep{NFW} following the formalism of \cite{LokasandMamon}; we chose $M_{\rm{NFW}}=10^{12} M_{\odot}$, $R_{\rm{vir}}=200$ kpc, and $c=10$ for the Milky Way. Recent studies have argued for a shorter thick-disk scale length \citep[e.g.;][]{Cheng2012,Bovy2012,Bensby2011}; however, this change would have a negligible effect on our results because of the comparatively low mass of the thick disk component.

The model can also be tuned for varying degrees of axisymmetry or triaxiality in the halo. For this study we use both spherical and triaxial models. For the triaxial model, we adopt the axis ratios $b/a = 0.99$ and $c/a = 0.72$ \citep{LawMajewski2010}. 

The Galactic Cartesian coordinate system used here is centered on the Galactic center; the $x$-axis points from the center toward the Sun (located at $x$ = 8.2 kpc \citep{Schoenrich2012}), the $y$-axis points along the direction of Galactic rotation, and the $z$-axis points toward the North Galactic Pole. To calculate velocity in this coordinate system, we convert radial velocity, distance, and proper motions to $U$, $V$, and $W$ in the Galactic coordinate system. Note that issues with astrometry in DR8, as explained in Section 3.5 of \citet{DR8} and the associated erratum \citep{DR8Error},  have been resolved for the DR9 astrometry used here. We choose the velocity of the local standard of rest to be 238 km s$^{-1}$, and the motion of the Sun with respect to that is $U = -13.8$ km s$^{-1}$, $V = 12.24$ km s$^{-1}$, and $W = 7.25$ km s$^{-1}$ \citep{Schoenrich2012,Schoenrich2010}.  This Galactic model is consistent with the measured proper motion of SgrA*, $6.379\pm0.026$ mas yr$^{-1}$ \citep{ReidBrunthaler2004}. Then, $U$, $V$, and $W$ are transformed into the Galactic Cartesian coordinate system, and we calculate the orbits backward in time for 1 Gyr using a fourth-order Runge$-$Kutta integrator. The choice of 1 Gyr is sufficient to discern the direction of origin while not being significantly influenced by a changing Galactic potential.

We examine the variation of each candidate's orbit given the errors described in Section 3.2. Figure 3 shows the 1$\sigma$ and 2$\sigma$ orbits for HVS 20, indicating that for some of the candidates the velocity errors are sufficiently large that the candidate itself may be bound. We find that the differences between orbits in the spherical versus the triaxial model are negligible for unbound orbits, since the stars have little time to respond to the halo potential. Therefore, for simplicity, when discussing unbound orbits we show only those in the spherical case. However, in instances when the orbit may be bound, as for HVS 20 in Figure 3, the halo shape definitely influences the candidate's trajectory, suggesting that marginally bound stars may help constrain halo triaxiality.

\begin{figure}[]
\centering
\includegraphics[width=0.5\textwidth]{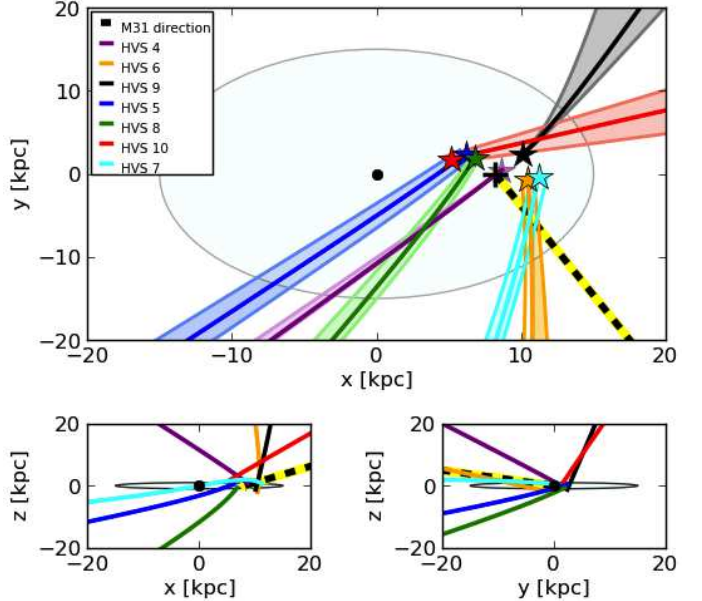} 
\caption[\, \,]{\footnotesize Orbits of the seven HVS candidates that are unbound with at least 98\% proability, over the past 1 Gyr. As in Figure 3, the black dots and plus signs represent the locations of the Galactic center and the Sun, respectively, while the pale blue ellipses provide a rough scale for the extent of the disk.  The shaded regions flanking the orbits of the same color represent the ``wedges" of possible orbits given the 1$\sigma$ velocity errors for the corresponding candidate. The like-colored stars mark the current positions of the candidates. None of the orbits plotted here intersect near the Galactic center, suggesting a different origin for these stars. 
In additional, if these stars had been traveling for 13 Gyr, they may have originated from as far as tens of megaparsecs away. The  dashed lines, highlighted in yellow for visibility, point toward M31's location 1 Gyr ago, assuming M31 proper motion, radial velocity, and distance from \cite{Sohn2012}. This interval was chosen to roughly coincide with the travel time of a hypervelocity star originating in M31. None of these HVS candidates seem to be coming from M31, which therefore is ruled out as a possible origin.}
\label{}
\end{figure}

\subsection{Origins}

As shown in Figure 4, the trajectories of these HVS candidates do not originate from the Galactic center, which would be expected if the stars were ejected by three-body interactions with the SMBH. Instead, they appear to be coming from all directions, which suggests that other ejection processes may be at play. 

We considered the SMBH at the center of M31 as a possible source \citep[][]{SherwinLoeb2008}, and given the velocities of the stars, we find the required flight time to reach the solar neighborhood would be approximately 1 Gyr. Figure 4 shows the orbits corresponding to the seven most unbound candidates, each with a 1$\sigma$ ``wedge" of possible orbits. It can be seen that the candidates could not have come from M31's SMBH position 1 Gyr ago (dashed lines). The orbits of the other 13 candidates are consistent with not arriving from M31. Therefore, the SMBH at the center of M31 is not responsible for ejecting these stars. However, this does not exclude other Galactic and extragalactic sources such as globular clusters, satellite galaxies, or the centers of distant galaxies within $\sim10$ Mpc.

\section{Discussion}
\label{discussion}

\subsection{Chemical Tracing}

Since it is more difficult to trace the past orbits of globular clusters and known satellite galaxies because of tidal stripping, shocks, and other mass-loss effects, we cannot say with certainty whether these stars could have originated in the Galactic disk, the bulge, or globular clusters. Another approach to determine whether these candidates belong to a particular population is to examine their chemical compositions.

We compared the metallicities of our candidates with the metallicity distribution functions (MDFs) of known globular clusters \citep{mcmaster}, the SEGUE G and K dwarf samples representative of the Milky Way disk population \citep{Schlesinger2012},  the Galactic bulge \citep{Sadler1996}, and by extension the bulge of M31, assuming a peak metallicity of $+$0.23 \citep{JacobyCiardullo99}. We also compared the metallicity distribution of our candidates with the MDF of the Galactic halo, although with a peak at $\rm{[Fe/H]} < -1$ \citep{An2013} it is clearly inconsistent with our candidates.

The MDFs for each population are shown in Figure 5. As perhaps expected, the metallicity of the HVS candidates is consistent with the G and K dwarf samples in the disk. Their metallicities are also largely consistent with the high-metallicity end of the globular cluster population and the low-metallicity end of the Galactic (and M31) bulge. Similarly, the stars' [$\alpha$/Fe]-values are broadly consistent with these stellar populations. Unfortunately, based on the information here, none of these populations can be decisively ruled out as a possible source, although it is clear that these HVSs do not originate from the metal-poor globular cluster system.


\begin{center}
\begin{deluxetable*}{@{}c@{}@{}cc@{}c@{}c@{}c@{}rrr@{}rrrcl}
\tablecolumns{14}
\tablecaption{Stellar and kinematic parameters for the 20 HVS candidates.}
\tablehead{ & & & & & \colhead{d} & & & & & & \colhead{\%} & \colhead{``Interloper} & \\
\colhead{HVS} & \colhead{IAU Name}  
& \colhead{r$_{0}$} & \colhead{[Fe/H]} &\colhead{[$\alpha$/Fe] \tablenotemark{a}} & \colhead{(kpc)} & \colhead{$v_{r}$\tablenotemark{b}} & \colhead{$v_{t}$\tablenotemark{c}} & \colhead{$v$\tablenotemark{d}} & \colhead{$v_{\rm min}$\tablenotemark{e}} & \colhead{$v_{\rm esc,Sph}$\tablenotemark{f}} &  \colhead{Bound} & \colhead{Likelihood"} & \colhead{Rating} }

\startdata
1	&	$J060306.77+825829.1$		
&	18.07 	&	$-$0.06 	&	0.10	&	3.70 	 &	$-$76.0 	&	56.1 		&	802.2	&	92.2		&	533.6	&	6.35	 	&	0.02		&	Clean\\
2	&	$J023433.42+262327.5$		
&	19.01 	&	$-$0.15 	&	0.09 	&	5.68  &	$-$25.6 	&	15.7 		&	628.6	&	290.0	&	517.3	&	7.43	 	&	0.18		&	Clean\\
3	&	$J160620.65+042451.5$		
&	19.01 	&	$-$0.91 	&	0.40 	&	4.06 	 &	 31.7		&	23.7 		&	641.8	&	195.1	&	588.9	&	34.88	&	0.15		&	Clean\\
4	&	$J113102.87+665751.1$		
&	16.15 	&	$-$0.83 	&	0.46 	&	1.04	 &	$-$54.9 	&	237.7 	&	1296.7	&	587.4	&	552.3	&	0.0	 	&	0.00		&	Reliable\\
5	&	$J185018.09+191236.1$		
&	18.16 	&	$-$0.34		&	0.19 	&	3.19	 &	 58.0 	&	61.5 		&	1086.8	&	378.9	&	576.5	&	0.04	 	&	0.00		&	Reliable\\
6	&	$J035429.27-061354.1$		
&	18.07 	&	$-$0.55		&	0.26 	&	3.13	 &	 80.2 	&	46.2 		&	916.3	&	286.6	&	534.5	&	0.07	 	&	0.01		&	Reliable\\
7	&	$J064337.13+291410.0$		
&	18.01 	&	$-$0.55 	&	0.35 	&	3.06	 &	 20.4 	&	38.1 		&	793.9	&	285.0	&	530.2	&	0.30	 	&	0.02		&	Reliable\\
8	&	$J202446.41+121813.4$ 	
&	17.74 	&	$-$0.65 	&	0.26 	&	2.48	 &	  6.26 	&	51.8 		&	769.1	&	376.3	&	570.3	&	1.01	 	&	0.03		&	Reliable\\
9	&	$J011933.45+384913.0$ 	
&	18.26 	&	$-$0.67 	&	0.22 	&	3.31	 &	$-$36.9 	&	65.5 		&	937.3	&	185.2	&	536.3	&	1.20	 	&	0.00		&	Reliable\\
10	&	$J172630.60+075544.0$ 	
&	18.46 	&	$-$0.67 	&	0.39 	&	3.82 	 &	  $-$2.2 	&	59.7 		&	992.9	&	233.5	&	591.0	&	1.34	 	&	0.00		&	Reliable\\
11	&	$J073542.35+164941.4$ 	
&	18.35 	&	$-$0.23 	&	0.12 	&	3.70	 &	 78.2 	&	28.8 		&	712.9	&	285.4	&	527.3	&	2.89	 	&	0.07		&	Reliable\\
12	&	$J025450.18+333158.4$ 	
&	18.25 	&	$-$0.70 	&	0.16 	&	3.14	 &	$-$62.4 	&	42.8 		&	731.4	&	265.1	&	532.9	&	3.77	 	&	0.05		&	Reliable\\
13	&	$J134427.80+282502.7$ 	
&	18.32 	&	$-$1.27 	&	0.44 	&	2.91	 &	   2.5 	&	44.0 		&	715.7	&	270.5	&	557.0	&	4.42		&	0.07		&	Reliable\\
14	&	$J225912.13+074356.5$ 	
&	18.76 	&	$-$0.56 	&	0.37 	&	4.60	 &	$-$97.8 	&	44.9 		&	840.7	&	121.8	&	550.0	&	5.86	 	&	0.01		&	Reliable\\
15	&	$J095816.39+005224.4$ 	
&	17.51 	&	$-$0.80 	&	0.28 	&	2.22	 &	   1.6 	&	59.2 		&	649.8	&	248.7	&	546.5	&	15.98	&	0.14		&	Reliable\\
16	&	$J074728.84+185520.4$ 	
&	17.81 	&	$-$0.24 	&	0.13 	&	3.26	 &	 43.9 	&	58.1 		&	672.8	&	55.3		&	530.7	&	19.70	&	0.11		&	Reliable\\
17	&	$J064257.02+371604.2$ 	
&	16.87 	&	$-$0.33 	&	0.21 	&	1.78	 &	   6.2 	&	49.1 		&	601.4	&	305.4	&	540.9	&	20.01	&	0.24		&	Reliable\\
18	&	$J165956.02+392414.9$ 	
&	19.22 	&	$-$1.14 	&	0.48 	&	4.35	 &   $-$205.1 	&	33.0 		&	649.1	&	170.0	&	562.3	&	21.30	&	0.14		&	Reliable\\
19	&	$J110815.19-155210.3$ 		
&	19.01 	&	$-$0.99 	&	0.35 	&	4.56	 &     131.2 	&	30.1 		&	622.7	&	162.0	&	545.8	&	23.69	&	0.19		&	Reliable\\
20	&	$J145132.12+003258.0$		
&	19.47	&	$-$0.59		&	0.12	&	5.88	 &	  88.0	&	16.5		&	606.7	&	193.1	&	579.8	&	43.24	&	0.23  	&	Reliable\\
\enddata
\tablenotetext{a}{Note there are large uncertainties in these measurements at the S/N of these candidates.}
\tablenotetext{b}{Radial velocity, in km s$^{-1}$, is taken straight from the SSPP without any corrections.}
\tablenotetext{c}{Total proper motion, in mas yr$^{-1}$, is calculated from the $\mu_{RA}$ and $\mu_{Dec}$ values listed in the CAS without any corrections.}
\tablenotetext{d}{This is the total velocity of the candidate, in km s$^{-1}$, after conversion to the Galactic Cartesian coordinate system described in \S 4.1, before any error consideration.}
\tablenotetext{e}{The value here is the minimum total velocity, in km s$^{-1}$, determined from a million realizations of proper motion, radial velocity, and distance errors drawn from their distributions described in Section 3.}
\tablenotetext{f}{The escape velocity, in km s$^{-1}$, at the star's position in a spherical potential.}
\end{deluxetable*}
\end{center}

\subsection{Alternative Origins}

As shown in Figure 4, none of the HVS candidates are coming from the Galactic center or from the direction of M31. The popular ejection mechanisms described inSection 1 involve a central SMBH and cannot explain these stars. The question where these stars originated, and how they gained such high velocities, remains.

One of the best-known hypervelocity mechanisms involves a binary system in the disk, in which a supernova explosion ejects the companion star \citep[e.g.;][]{Blaauw1961}. There are many lesser known hypervelocity ejection mechanisms as well. For example, multibody ejections from the dense central regions of globular clusters \citep[e.g.;][]{Poveda1967} including globular clusters that may have dissipated over the lifetime of the Galaxy \citep[e.g.;][]{Gnedin1997,Chernoff1990,McLaughlin2008} may boost a star to hypervelocities. In addition, there could be a three-body interaction involving an intermediate-mass black hole or otherwise very massive star \citep[e.g.;][]{Gvaramadze2009}. A final hypervelocity ejection mechanism involving a stellar dynamical process could be the partial tidal disruption of a single star around a SMBH \citep[][]{Manukian2013}.

Furthermore, three-body interactions between galaxies, such as M31 \citep[e.g.;][]{Caldwell2010}  and the Large and Small Magellanic Clouds \citep[e.g.;][]{Chandar2010}, have been suggested as possible hypervelocity ejection mechanisms, although we have already ruled out M31 specifically. HVSs may also receive an energy boost during the tidal stripping process as long streams are stripped from an accreted satellite \citep[e.g.;][]{Abadi2009,Caldwell2010,Piffl2011,Fouquet2012,King2012}. 

\begin{figure}[]
\centering
\includegraphics[width=0.5\textwidth]{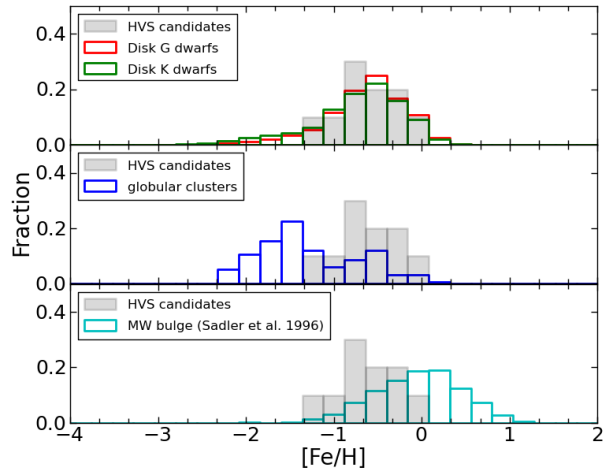}
\caption[\, \,]{\footnotesize Normalized metallicity distribution functions of our candidates (shaded), compared with G dwarfs (red), K dwarfs (green), globular clusters (blue), and the Galactic bulge (cyan).}
\label{}
\end{figure}

\subsection{Follow-up Analysis}

A significant fraction of the candidates failed the {\tt dist22 $>$ 7} requirement, meaning that photometric blending from a near neighbor may have affected the proper-motion determination. A larger number of the candidates suffer from too few detections in the SDSS+USNO-B catalog. Confirming these candidates as HVSs would require additional astrometric analysis in order to verify their proper motions; the {\it Hubble Space Telescope} Fine Guidance Sensor may be appropriate.  

There is also the possibility that these candidates are unresolved spectroscopic binaries, which could imprint a large radial velocity signal. Future, higher resolution spectroscopic observations could easily decide this issue and would also allow a more detailed chemical analysis to shed light on their origins.

\subsection{Constraints on the Initial Mass Function}

The fact that we find no low-mass HVSs coming from the Galactic center continues to pose a problem for a universal initial mass function and an unbiased binary ejection mechanism. If we simply assume a Salpeter initial mass function and a mass-blind dynamical process, we would naively expect roughly 150 HVSs in the 0.6$-$1.2 solar mass range in our sample, compared with the 14 known 3$-$4 M$_{\odot}$ HVSs \citep{Brown2009}. Either the initial mass function near the Galactic center is top-heavy or the process acting at the Galactic center ejects  over 10 times more high-mass stars than low-mass ones.  There is tentative evidence from the Arches and other young star clusters at the Galactic center that the initial mass function is top-heavy, with a slope of about $-$1.6 \citep{Figer1999}, although this is a matter of debate. If we adopt this slope for our initial mass function, then we still should have observed roughly 40 HVSs with spectral types G and K, which would require an ejection mechanism that favors massive stars by more than factor of  3.

Our constraints on the initial mass function are consistent with the findings of \cite{Kollmeier2010}, who searched for metal-rich F/G halo HVSs. This earlier study placed stricter limits on the ejection mechanism, however, because F/G stars would be expected to accumulate in the halo over their main-sequence lifetimes, while our sample probes only stars passing through the solar neighborhood; stars ejected from the Galactic center through stellar binary disruption, for example, would reach and pass through our sample in mere tens of millions of years. Our results are also consistent with the constraints from \citet{Zhang2013}, who considered the S stars at the Galactic center to be the captured companions of binary star tidal disruption, a process that ejects the second star.

\newpage
\section{Summary and Conclusions}

We report a set of 20 hypervelocity candidates from the SEGUE G and K dwarf sample. These candidates have velocities greatly exceeding the escape velocity at their respective positions in the Galaxy, albeit with large proper-motion errors. Monte Carlo estimates of the position and kinematics of these stars show that seven of the 20 exceed the escape velocity at their respective locations within the Galaxy with at least 98\% probability and that each candidate's interloper likelihood is less than 25\%.

Surprisingly, an orbit analysis indicates that none were ejected from the Galactic center. The confirmation of these candidates as HVSs argues for a more careful exploration of alternative ejection mechanisms such as interactions within globular clusters, dwarf galaxies, or tidal tails, as well as ejections from supernovae in the Galactic disk.

If these stars are truly hypervelocity, their spectra could already contain clues to their origin. For example, abundance patterns indicative of supernova contamination would confirm or rule out a candidate's having been ejected from a high-mass binary system \citep{Przybilla2008}.

One remaining question is why these stars were not identified in previous HVS campaigns. A possibility is that prior searches focused on extreme radial velocities \citep[e.g.;][]{Brown2005}. While the radial velocities of our candidates are relatively modest, it is the addition of proper motions that boosts these stars into hypervelocity candidacy. Naturally, our sample also explored a cooler spectral type than previous work. Future surveys may be more successful in identifying HVSs with both radial velocity and proper-motion measurements.

We are expanding our search for HVS candidates to the entirety of the SDSS DR9 sample in order to include all spectral types. Analysis of any additional candidates identified in this search, as well as follow-up, is deferred to a future paper.

\section{Acknowledgments}

We acknowledge David Weinberg for useful suggestions on our manuscript.
 
LEP was supported by the Graduate Assistance in Areas of National Need program. KHB acknowledges support from NSF award AST 0847696, and the Aspen Center for Physics. KJS acknowledges support from NSF grants AST 0807997 and AST 0607482. TCB acknowledges partial funding from NSF Physics Frontier Center grants PHY 0216783 and PHY 0822648 to the Joint Institute for Nuclear Astrophysics (JINA).

Funding for SDSS-III has been provided by the Alfred P. Sloan Foundation, the Participating Institutions, the National Science Foundation, and the U.S. Department of Energy Office of Science. The SDSS-III web site is http://www.sdss3.org/.

SDSS-III is managed by the Astrophysical Research Consortium for the Participating Institutions of the SDSS-III Collaboration including the University of Arizona, the Brazilian Participation Group, Brookhaven National Laboratory, University of Cambridge, Carnegie Mellon University, University of Florida, the French Participation Group, the German Participation Group, Harvard University, the Instituto de Astrofisica de Canarias, the Michigan State/Notre Dame/JINA Participation Group, Johns Hopkins University, Lawrence Berkeley National Laboratory, Max Planck Institute for Astrophysics, Max Planck Institute for Extraterrestrial Physics, New Mexico State University, New York University, Ohio State University, Pennsylvania State University, University of Portsmouth, Princeton University, the Spanish Participation Group, University of Tokyo, University of Utah, Vanderbilt University, University of Virginia, University of Washington, and Yale University.

\bibliographystyle{apj}

\end{document}